\documentclass{iau} 
\usepackage{graphicx}

\title[UFO, accretion and ejection in AGN] 
{Ultra fast outflows, and their connection to accretion and ejection processes in AGNs}

\author[A.L. Longinotti et al.]   
{Anna Lia Longinotti 
 }

\affiliation{CONACyT--Instituto Nacional de Astrof{\'i}sica, {\'O}ptica y Electr{\'o}nica \\ Luis E. Erro 1, Sta Mar{\'i}a Tonantzintla, Puebla, C.P. 72840, Mexico 
 \\ email: {\tt annalia@inaoep.mx} \\[\affilskip]
 }

\pubyear{2018}
\volume{IAUS342}  
\jname{Perseus in Sicily: from black hole to cluster outskirts}
\editors{Keiichi Asada, Elisabete de Gouveia dal Pino, \\ Hiroshi Nagai, Rodrigo Nemmen, Marcello Giroletti}
\begin{document}

\maketitle

\begin{abstract}
The growing evidence for energy-conserving outflows in powerful and luminous AGN supports the idea that high-velocity winds launched from the accretion disc evolve systematically after undergoing a shock with the ambient medium and that they are capable to expel enough mass and energy so as to produce feedback. 
This talk will give an overview of recent results on AGN ultra fast outflows, with focus on
grating X-ray spectra of bright sources.
I will review how UFO work, their observational properties and their relation with AGN outflows in other bands, what is their impact on the host galaxies and their role in feedback processes. 
\keywords{galaxies: active, hydrodynamics, atomic processes}
\end{abstract}

\firstsection 
\section{Introduction}

Feedback from Active Galactic Nuclei (AGN) is generally thought to be an important  ingredient for galaxies evolution.
After large amount of gas is accreted during the earliest stage of a quasar life time, the accumulated energy can be released  via ejection of powerful outflows driven by the AGN. 
If the outflow is as strong as 0.5--5\% of the Eddington luminosity of the AGN, it has a profound impact on the development of the host galaxy itself.  The effect of these winds is to eventually expel the gas that would otherwise be available for forming new stars in the host galaxy therefore providing an effective mechanism of quenching star formation. 
It is in this sense that we refer to AGN feedback as a mechanism able to  regulate the growth of the galaxy and the growth of the central black hole as well (Di Matteo et al. 2005, Hopkins et al. 2010).

A widely accepted scenario for explaining AGN feedback postulates  that a fast wind observable in the X-ray band is launched at accretion disk scale (Faucher-Gigu{\`e}re \& Quataert 2012). This highly ionized X-ray gas is currently observed in the form of Ultra Fast Outflows in some AGN spectra (Tombesi et al. 2012, Gofford et al. 2013). While traveling outward, the impact of the wind with the ISM (inter-stellar medium)  gives rise to shock processes (King 2010).  After shocking with the gas, deceleration and cooling processes lead to the production of a  slower outflow with  less ionized lines observable in the optical band and to the formation of a bubble of hot, tenuous gas (e.g. Zubovas \& King 2012).  As a result of the cooling, the presence of molecular gas outflowing at a much lower velocity is expected. This latest phase is frequently observed in several ULIRGS and Quasars (Cicone et al. 2014, Feruglio et al. 2010, 2015).  

To date, only two  cases of ULIRGS are reported where the observed X-ray and molecular phases of the outflow are physically related, IRAS~F11119+3257 (Tombesi et al. 2015) and Mrk~231 (Feruglio et al. 2015). Both results remarkably fit in with the prediction of the energy-conserving outflow model outlined above. \\

\begin{figure}[b]
\begin{center}
 \includegraphics[width=12cm]{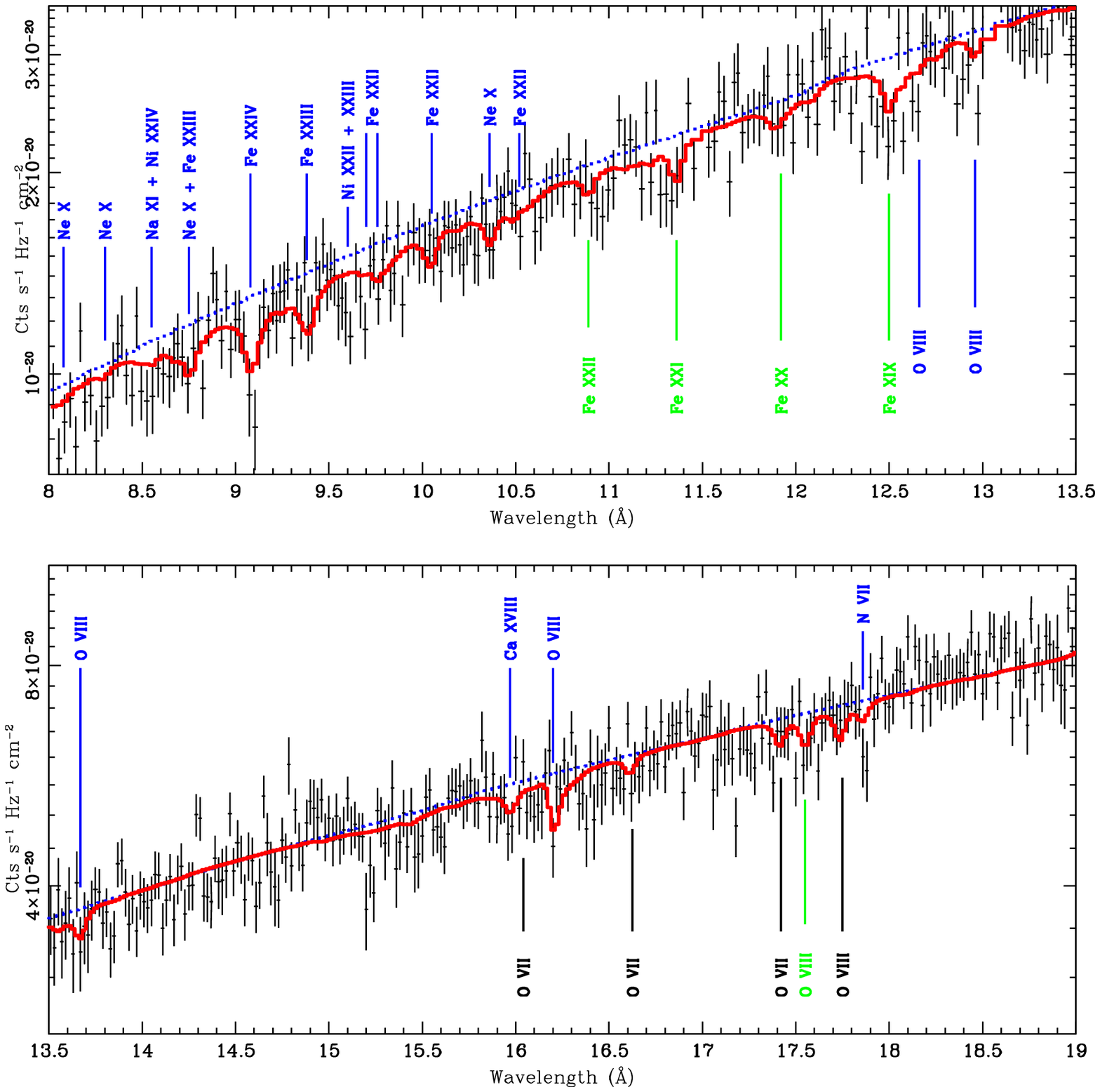} 
 \caption{{\it XMM-Newton}-RGS spectrum of the NLSy1 Galaxy Mrk~1044 fitted by four components of ultra fast outflows (Krongold et al. submitted). The fastest component  (blue labels) of this system reaches an outflow velocity of $\sim$48,000~km~s$^{-1}$  and column density of N$_H$$\sim$10$^{23.32}$~cm$^{-2}$, while the other three are outflowing at 25,000~km~s$^{-1}$. }
   \label{fig_1044}
\end{center}
\end{figure}

This fascinating picture, among several other variables, relies on the existence and the properties of the nuclear wind, the so-called ``Ultra Fast Outflow". 
The X-ray spectra of some AGNs  show  signature of gas outflowing at high speed ($v \ge 0.1$~c). This gas is so highly ionized by the nuclear radiation that the only dominant ions left are He-like and  Hydrogen-like ionic species. These systems were christened as ``Ultra-Fast Outflows" (UFOs)  (Tombesi et al. 2010) and at the beginning they were observed mainly in the Fe K band at E$\ge$~7~keV.
Several papers reported on UFOs hosted in individual AGNs (Pounds et al. 2003, 2011, 2014, Chartas et al. 2009, Lanzuisi et al. 2012, Nardini et al. 2015), and statistical studies show that they are detected in 30-40\% of nearby AGN (Tombesi et al. 2010, 2012, Gofford et al. 2013).  The approximate ranges of mass outflow rate (0.01-1M$_{\odot}$~yr$^{-1}$) and kinetic energy (10$^{42-45}$ erg~s$^{-1}$) are in good agreement with theoretical predictions for black hole winds (King 2010). 
   
Two launching mechanisms are envisaged for the production of such rapid outflows, and both locate the region of action in the accretion disk of the AGN. One way to produce a fast outflow is through radiation driving (Proga and Kallman 2004, Sim et al. 2010), and another one is via magnetic fields in radio-loud sources (``magnetically driven outflows", Fukumura et al. 2015).

\section{Ultra Fast Outflows in Narrow Line Seyfert 1 Galaxies}
\begin{figure}[t]
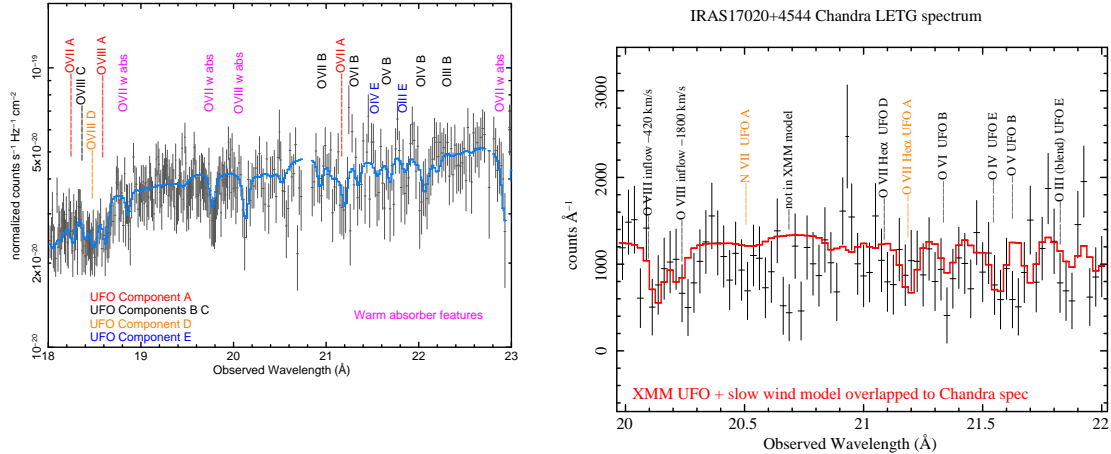

 \vspace*{1.0 cm}
\includegraphics[width=5.cm,angle=-90]{fig2a.ps} 
 \includegraphics[width=6.cm, angle=-90]{fig2b.ps} 
 \caption{The system of outflowing and inflowing gas in the NLSy1 Galaxy IRAS17020+4544. {\it Left}: Portion of the {\it XMM-Newton}-RGS spectrum (observed in 2014) mostly affected by the fast wind absorption. Letters indicate the velocity component of the outflow as described in L15.
 The label ``Warm absorber"  marks lines that are part of the slow ionized outflow. 
  {\it Right}: Close-up of the {\it Chandra}-LETG spectrum of the source (observed in 2017) with the XMM model overlapped  (without fitting) in the spectral region where ``slow" inflow and fast outflow are visible (Longinotti et al. in prep.). Labels refer to absorption features identified in the {\it XMM-Newton}-RGS spectrum (L15 and Sanfrutos et al. submitted).}
   \label{fig_letg}
\end{figure}
In the very recent years, the discovery of fast outflows in the soft X-ray grating spectra of bright Seyfert Galaxies has opened a complementary path to explore the properties of AGN fast winds. Compared to the results based on CCD spectra, the higher detail provided by grating spectroscopy has revealed that soft X-ray fast winds can be made by multiple components of distinct outflowing velocity, ionization state and column density and that they can be massive enough to produce feedback in the host galaxy, e.g. IRAS~17020+4544 (Longinotti et al. 2015, L15 throughout) and Mrk~1044 (see Fig~\ref{fig_1044}, Krongold et al. submitted).  When luminosity variations takes place, the combination of timing and spectroscopy techniques shows that fast X-ray winds respond to flux variations revealing a tight connection between continuum photons emitted in the very inner accretion disc with highly ionized outflowing gas (IRAS~13224-3809 Parker et al. 2017; PDS~456 Matzeu et al. 2017). With the exception of the luminous QSO PDS~456 (Reeves et al. 2016), if we consider the other few cases of fast winds observed in gratings spectra  (PG1211+143, Reeves et al.  2018; Akn~564, Gupta et al. 2013), it is clear that  the sources where this phenomenon is detected share the same classification: they are all Narrow Line Seyfert 1 (NLSy1). 

This AGN class represents an extreme form of Seyfert activity that is manifested in their peculiar continuum and emission-line properties in almost all bands (see Gallo  2006), their small black hole masses (M$_{BH}$$\sim$ 10$^{6-7}$ M${\odot}$), high Eddington ratios and often, high degree of X-ray variability (Komossa \& Xu 2007). 

The recent findings of fast X-ray winds in an increasing number of NLSy1 seem to hint to a common behavior of these sources with objects of much higher luminosity (e.g. Chartas et al. 2002, Tombesi et al. 2015, Nardini et al. 2015). Increasing evidence that the velocity of the wind correlates with the source luminosity (i.e. winds are faster at higher luminosities,  Pinto et al. 2018)  strongly favours  radiation driving as the launching mechanism for the outflows, leading  therefore to the hypothesis that a high accretion rate is the likely driver of fast winds  (Matzeu et al. 2017).

\section{Stratification of the fast outflow: a shocked outflow origin?}

We now focus on the peculiar case of the NLSy1 IRAS17020+4544. This source presents the simultaneous presence of a stratified fast outflow (L15) and a multi-layered slower absorber. 
Fine quality multi-epoch spectroscopic information  available from two {\it XMM-Newton} observations (obtained in the years 2004 and 2014) and a long look with {\it Chandra} LETG (obtained in 2017) confirm a surprisingly steady continuum flux and a complex pattern of absorption features (see Fig.~\ref{fig_letg}).
While no significant variations are observed in the properties of the fast wind along the 10~ years elapsed in between the two {\it XMM-Newton} observations, the peculiar evolution of the 4 components of the slow absorber  (Sanfrutos et al. submitted) reveals gas flowing inward (v$_{inflow}$$\sim$1800-3000~km~s$^{-1}$) and outward (v$_{outflow}$$\sim$2000-3400~km~s$^{-1}$), thus suggesting a different nature of this wind compared to ``standard'' warm absorber.    

\begin{figure}[t]
 \vspace*{1.0 cm}
 \includegraphics[width=8cm]{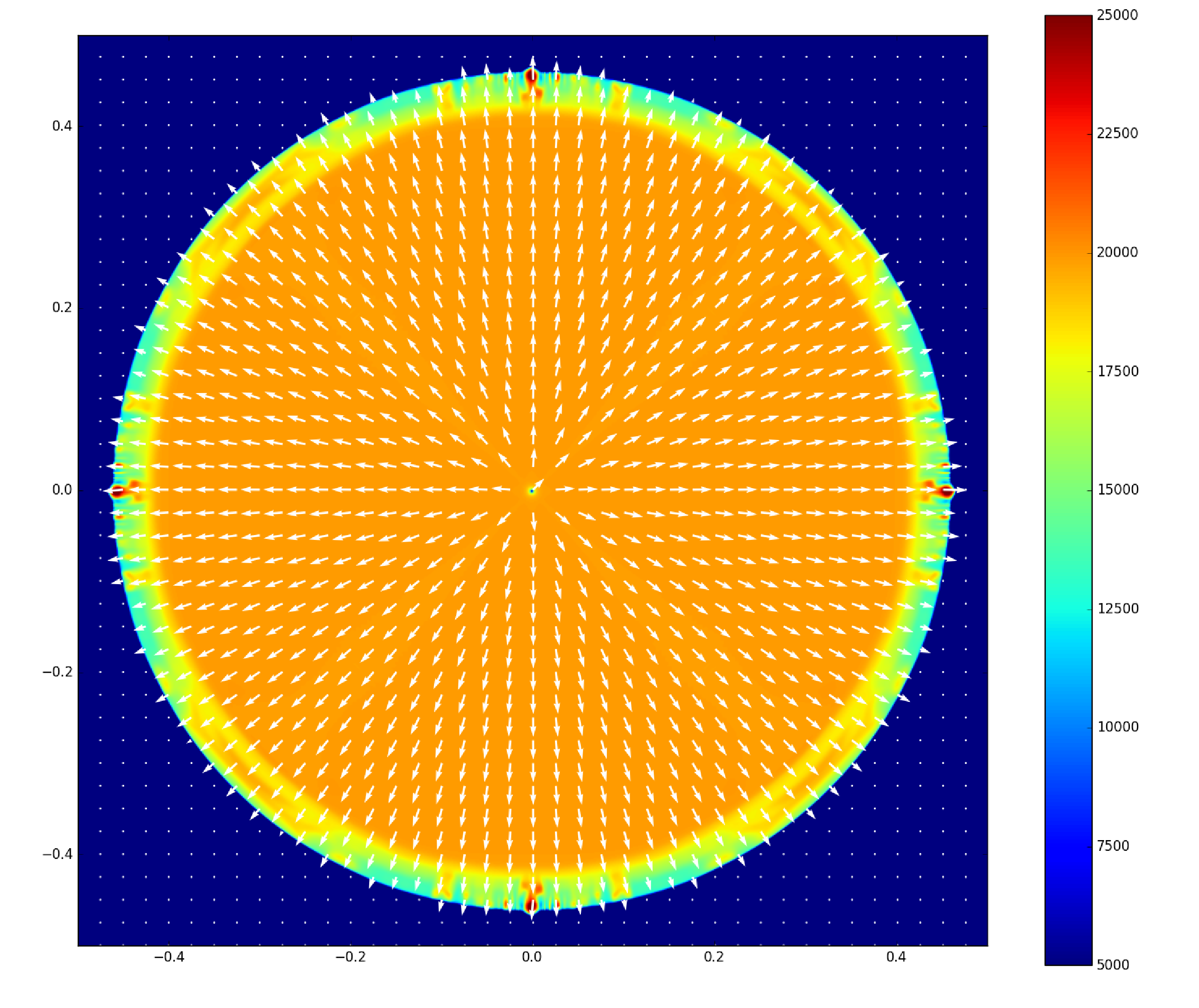} 
  \includegraphics[width=8cm]{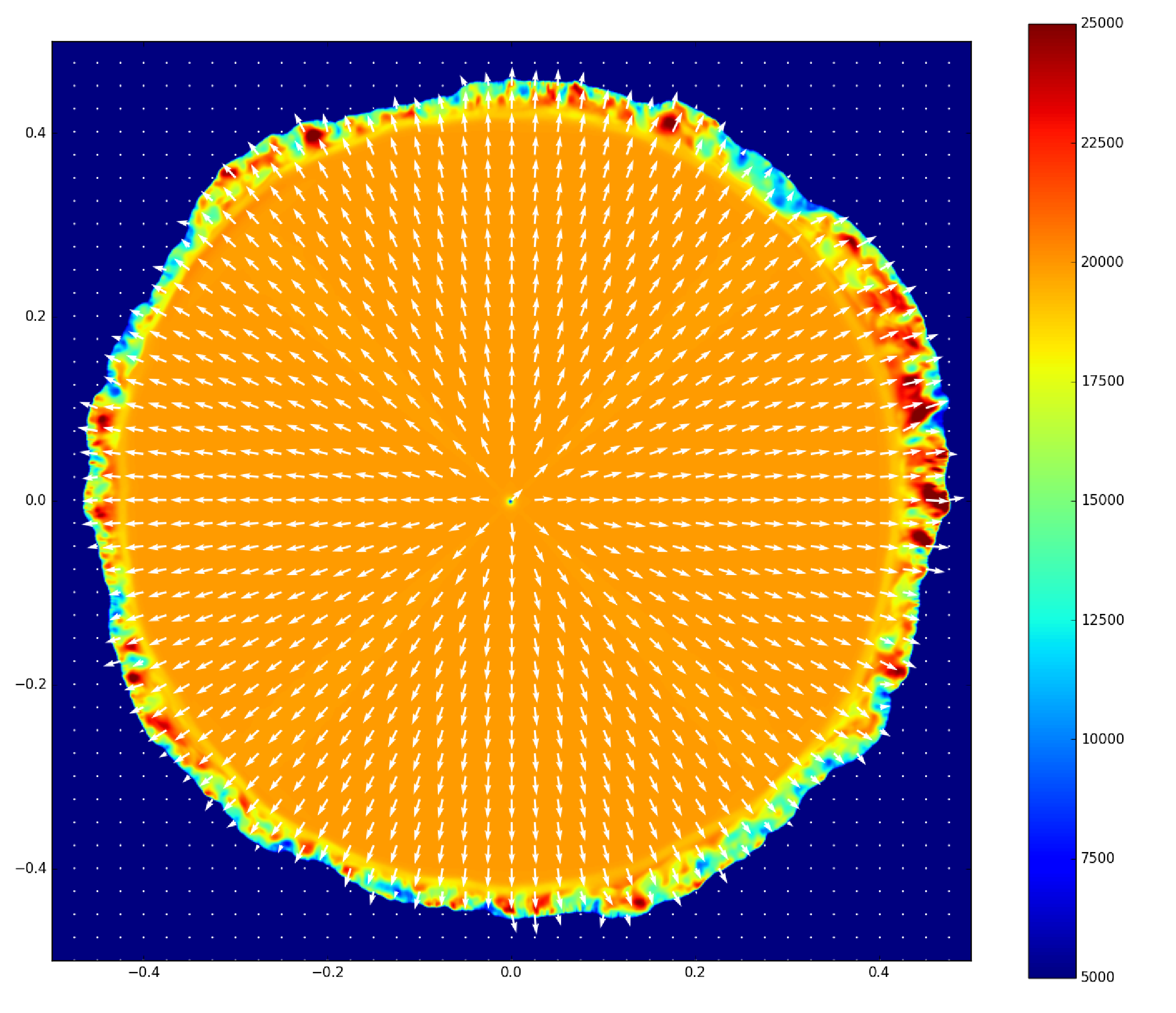} 
 \caption{Toy model showing the appearance of a shocked outflow affected by instabilities. The color scale maps the velocity of the ultra fast outflow ejected at  20,000~km~s$^{-1}$ that is  pushing the shock outward ({\it Left}). When instabilities are introduced ({\it Right}), sections of gas with higher velocities start to appear at the border of the shock (simulations to be included in Longinotti et al. in prep.)}
   \label{fig_simul}
\end{figure}

To explain this complex pattern of absorbers we postulate that the coincidence of the fast outflow with slower winds moving in opposite directions may be explained in terms of a ``shocked outflow''.
This model predicts that an initial fast outflow  radiatively launched at accretion disc scale with outflow velocity  $v_{\rm out}$~$\ge$ 10$^4$~km~s$^{-1}$ shocks with the ambient medium (see King \& Pounds 2015 for a review). 
  The two shock fronts (reverse and forward)  produced by the impact of the wind with gas at an escape velocity lower than the outflow velocity, are separated by a contact discontinuity and whereas the shocked ambient gas could decelerate to velocity of the order of 10$^2$~km~s$^{-1}$, the wind shock (forward) maintains its high velocity while entraining the ambient gas and pushing it further out (Faucher-Gigu{\`e}re \& Quataert 2012).

  Analogously to  Supernova Remnants (Velazquez et al. 1998),  fluid instabilities (\eg\ Rayleigh-Taylor) are likely to develop at the discontinuity between the two shock fronts due to the difference between  the densities of the impacting wind and of the impacted medium. A condition for the Rayleigh-Taylor instability to grow is that the mass of the ISM that is pushed by the discontinuity is higher than the mass of the ejecta (Velazquez et al. 1998), which undergoes to a deceleration process that is able to trigger instabilities in the fluid (see Fig.~\ref{fig_simul}).  Such instabilities would easily alter the dynamics of the shocked outflow and they may give rise to a re-distribution of the overall velocity field in which slower (and faster) components of the wind may simultaneously cross our line of sight.  The effect of the turbulence would then provide a replenishment mechanism that continuously supply new sections/blobs of gas that might  as well fall backwards with an opposite direction with respect to the bulk of the outflow.

We have developed a toy model for simulating this behavior and testing the viability of this idea. 
The simulation is based on the GUACHO code (Esquivel \& Raga 2013) and a preliminary output is displayed in Fig.~\ref{fig_simul}.  An expanding shock is pushed within a turbulent medium by an inner wind with v$_{out}$=20,000 km~s$^{-1}$. The expansion time is set to 20~yr , the mass of the central object is 10$^6$ M$\odot$ (i.e. the mass of IRAS17020+4544 and in general of the order of black hole masses in NLSy1 galaxies). 

  Fig.~\ref{fig_simul} shows the pattern of velocities developed at the shock front with and without instability. The right panel shows that ``plumes" or ``fingers" of gas with different velocities are formed as a result of the introduction of instabilities process in the expansion of the shocked outflow. Depending on which ``fingers" are intercepted by  our line of sight, it is evident  that the resulting absorption line spectrum is likely to show a stratified wind consistent with the spectrum observed in IRAS17020+4544 (see Fig.~\ref{fig_letg}). 
  This scenario can easily apply to other NLSy1 sources with the same observational properties.

\section{A truly multi-phase outflow in IRAS17020+4544?}
\begin{figure}[t]
 \vspace*{1.0 cm}
 \includegraphics[width=7cm]{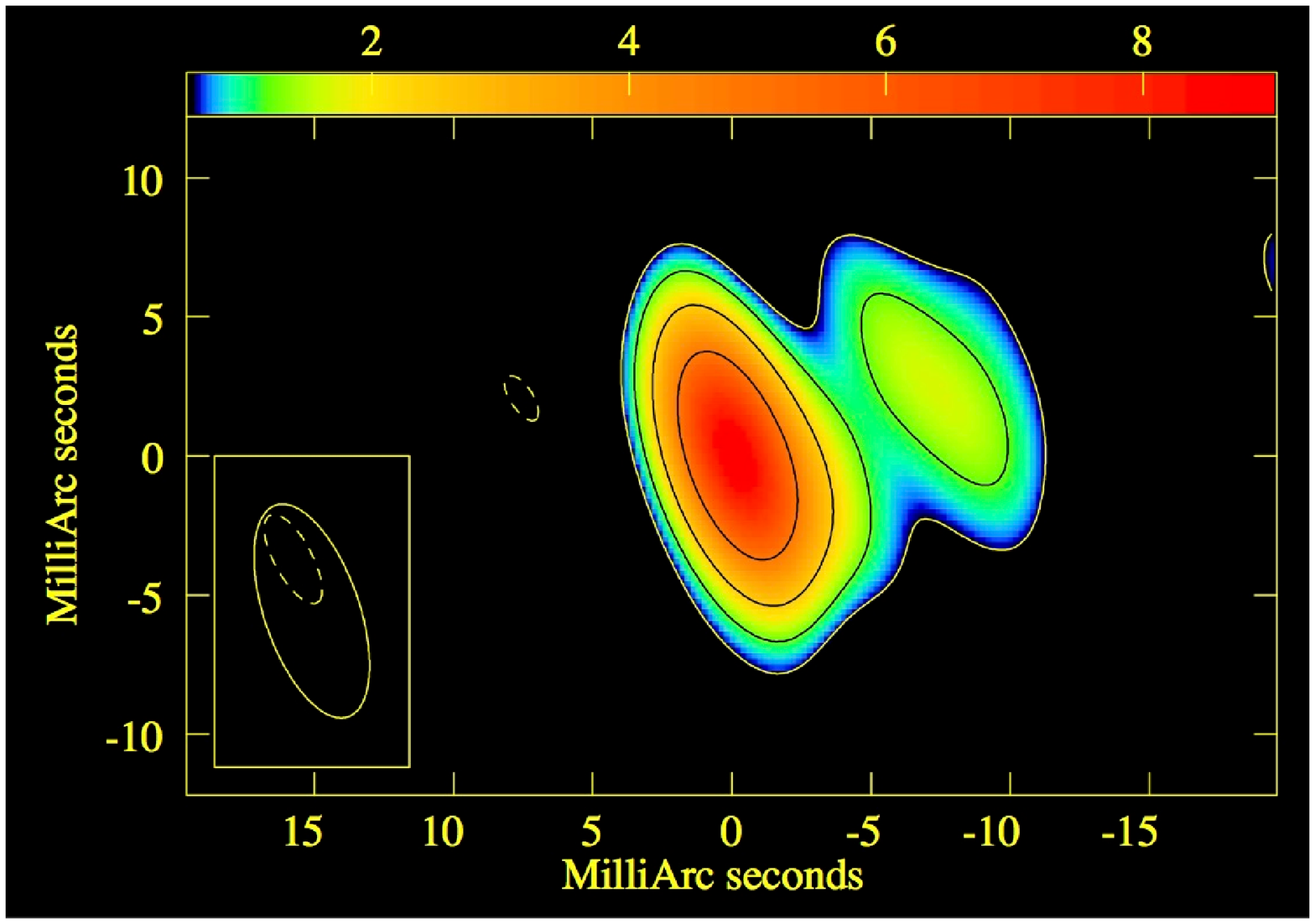} 
 \includegraphics[width=7.5cm]{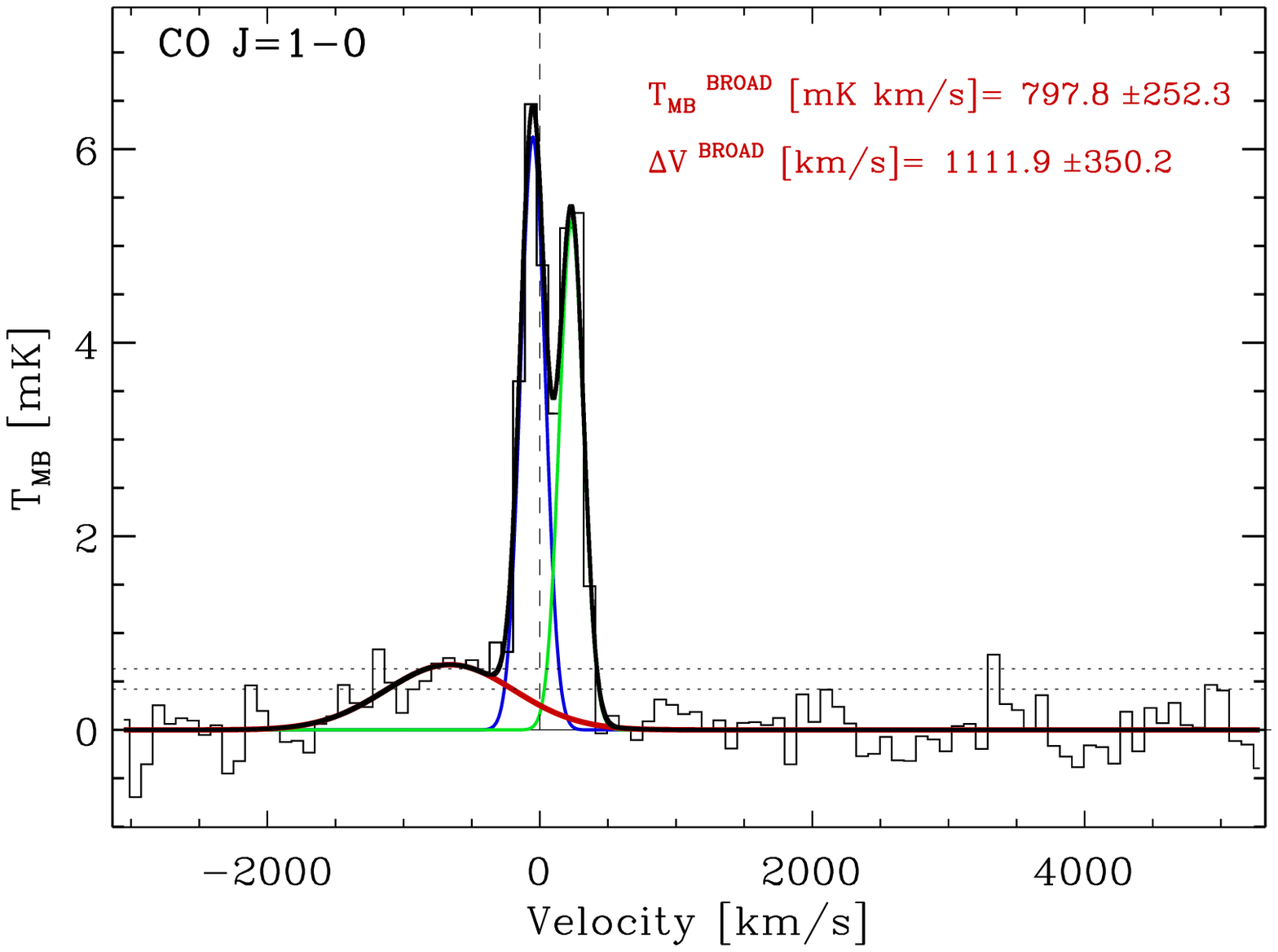} 
 \caption{{\it  Left}: Simultaneous 5-8 GHz spectral index VLBA image of IRAS17020+4544 observed in 2014. Contours show total intensity at 8 GHz and colours show the spectral index values (Giroletti et al. 2017). {\it Right}: Spectrum of the CO(1-0) line in IRAS17020+4544 obtained with the LMT telescope in 2017. The double peaked structure is fitted by two narrow Gaussian components (blue and green), and the line wings are fitted by broad Gaussian line (red line). 
 The molecular gas mass is estimated assuming a CO-to-H2 conversion factor appropriate for ULIRGs (Solomon et al. 1997). The molecular gas masses estimated in each narrow component are very similar, which is compatible with the presence of a ring of molecular material in the galaxy, while the blue wing of the line is consistent with the presence of molecular gas outflowing at  $\sim$660 km~s$^{-1}$  (Longinotti et al. submitted). }
   \label{fig_radio}
\end{figure}

Further hints to the presence of a shocked outflow in this source come from radio observations (see Fig.~\ref{fig_radio}) that have revealed an elongated structure on a scale of 10~pc in VLBI images (Giroletti et al. 2017).  
The appealing possibility that such compact jet, which is produced by synchrotron emission, may represent the signature left by the shock of the inner X-ray outflow with the ambient gas has been postulated by several models of galactic outflows (e.g. Zakamska \& Greene 2014; Nims et al. 2015) in an attempt to link synchrotron emission in radio-quiet sources with galaxy scale molecular outflows. 

IRAS17020+4544 is hosted by a barred spiral galaxy with  IR luminosity typical of LIRG (L$_{FIR}$= 1.05$\times$10$^{11}$$L_{\odot}$) therefore likely to be rich in molecular gas. 
We have obtained milimetric observations with the Large Millimetric Telescope (LMT,  Hughes et al. 2010) and found tentative evidence for a  galaxy scale outflow traced by CO gas (Fig.~\ref{fig_radio}, Longinotti et al. submitted).

Assuming a dynamical timescale  of few$\times$10$^{6}$ years for the outflow to propagate out of the nucleus at the observed bulk velocity of $\sim$600~km~s$^{-1}$, the spatial scale of the wind is approximately $\sim$0.5-3~kpc. 
The  CO mass in the wind is of the order of 10$^8$~M$\odot$  and following Feruglio et al. 2015, we find an approximate estimate of the molecular outflow rate of $\sim$100~$M_{\odot}$yr$^{-1}$, which is consistent with recent findings reported for Active Galaxies with comparable CO properties (Cicone et al. 2014)

The relative proportion of momentum load for the X-ray and molecular outflows in IRAS17020+4544 would provide an additional evidence for the existence of energy conserving winds  that propagate through the galaxy after undergoing a momentum boost, supporting the feedback scenario as proposed in earlier works (Feruglio et al. 2015). 

Admittedly, these numbers suffer from the large uncertainties in the estimates of  the mass outflow rate  and of the wind spatial extent, therefore we are currently awaiting the outcome of an already  performed interferometry observation to constrain the size of the  molecular outflow region and its physical properties.

\section{Acknowledgments}
The author wishes to thank the organizers of the IAU symposium 342  ``Perseus in Sicily: from black hole to cluster outskirts". Financial support from the  ``Coordinaci{\'o}n de Astrof{\'i}sica" of INAOE Institute is gratefully acknowledged. The author thanks the many colleagues who, through their constant streaming of input and discussion, have actively collaborated to the work presented at this conference: Yair Krongold, Pablo Velazquez, Mario Sanfrutos at UNAM (Mexico City), Olga Vega and Vahram Chavushyan at INAOE (Puebla), Matteo Guainazzi (ESA-ESTEC, Leiden), Marcello Giroletti (INAF-IRA, Bologna), Francesca Panessa (INAF-IAPS, Roma), Elisa Costantini (SRON, Utrecht).

\end{document}